\documentclass{article}

\usepackage{arxiv}

\usepackage[utf8]{inputenc} 
\usepackage[T1]{fontenc}    
\usepackage{hyperref}       
\usepackage{url}            
\usepackage{booktabs}       
\usepackage{amsfonts}       
\usepackage{nicefrac}       
\usepackage{microtype}      
\usepackage{lipsum}
\usepackage{csquotes}
\usepackage{multirow}
\usepackage[table]{xcolor}
\usepackage{makecell}
\usepackage{graphicx}
\graphicspath{ {./images/} }
\usepackage{subcaption}

\title{Topology and dynamics of narratives on Brexit propagated by UK press during 2016 and 2017}

\author{
  Jorge Louçã\thanks{Corresponding author.} \\
  ISTAR-IUL, cacifo 296\\
  Instituto Universitário de Lisboa (ISCTE-IUL)\\
  Av. das Forças Armadas, 1649-026 Lisboa, Portugal \\
  \texttt{Jorge.L@iscte-iul.pt}
   \And
 António Fonseca \\
  ISTAR-IUL, cacifo 296\\
  Instituto Universitário de Lisboa (ISCTE-IUL)\\
  Av. das Forças Armadas, 1649-026 Lisboa, Portugal \\
  \texttt{ajffa@iscte-iul.pt} \\
}

\begin{document}
\maketitle

\begin{abstract}
This article identifies and characterises political narratives regarding Europe and broadcasted in UK press during 2016 and 2017. A new theoretical and operational framework is proposed for typifying discourse narratives propagated in the public opinion space, based on the social constructivism and structural linguistics approaches, and the mathematical theory of hypernetworks, where elementary units are aggregated into high-level entities. In this line of thought, a narrative is understood as a social construct where a related and coherent aggregate of terms within public discourse is repeated and propagated on media until it can be identified as a communication pattern, embodying meaning in a way that provides individuals some interpretation of their world. An inclusive methodology, with state-of-the-art technologies on natural language processing and network theory, implements this concept of narrative. A corpus from the Observatorium database, including articles from six UK newspapers and incorporating far-right, right-wing, and left-wing narratives, is analysed. The research revealed clear distinctions between narratives along the political spectrum. In 2016 far-right was particularly focused on emigration and refugees. Namely, during the referendum campaign, Europe was related to attacks on women and children, sexual offences, and terrorism. Right-wing was manly focused on internal politics, while left-wing was remarkably mentioning a diversity of non-political topics, such as sports, side by side with economics. During 2017, in general terrorism was less mentioned, and negotiations with EU, namely regarding economics, finance, and Ireland, became central.
\end{abstract}

\keywords{brexit \and narratives \and press, social constructivism \and structural linguistics \and hypernetworks \and natural language processing \and political discourse \and narrative topology \and narrative dynamics}

\section{Introduction}

United Kingdom voted to leave the European Union in a referendum on the 23\textsuperscript{rd} of June, 2016. The leave procedure was initiated on the 29\textsuperscript{th} of March, 2017, by the activation of article 50 on the European Union Treaty. The whole process of public political decision-making, and then negotiation, has been named “Brexit”, a mix of “Great Britain” and “exit”.

The debate concerning the presence of United Kingdom in the European economic and political institutions exists since its early admission to the Economic European Community. The first referendum on this subject took place in June 1975, when it was decided for remaining. However, since then the opposition to the European integration kept being active within UK political debate. The Eurosceptic ideas get stronger inside several political wings, from left-wing to right-wing. Recently far-right arguments had a real impact in society. On the 20th of January 2016, a new referendum was announced for deciding to remain or to leave the European Union. Since then, this debate became central in UK society. The year of 2016 was marked by the “Leave” vs. “Remain” campaigns. The result of the referendum, surprising for some, leaded, during final 2016 and the whole year of 2017, to a heated debate on the consequences of leaving, and then on negotiations with European Union on how to achieve this political process, which affects economic, financial, and demographic domains~\cite{1}~\cite{2}~\cite{3}.

Crossing arguments for “Leave” or “Remain” within all political currents have been particularly active in social networks and UK press~\cite{4}. A diversity of studies, both qualitative and quantitative, was dedicated to analysing this debate in media~\cite{5}~\cite{6}. However, we argue that a precise quantitative analysis is still missing, allowing to characterise arguments and synthesize the main lines of political thought that have been disseminated in public opinion.

\subsection{Goal}

The goal of this research is to identify and to characterise the political narratives broadcasted in UK press during 2016 and 2017. We start by noticing the need for defining a new theoretical and operational framework capable of explaining narratives propagated in the public opinion space, and therefore able to describe the Brexit debate dynamics.

Our main hypothesis is that narratives in public opinion can be characterised through automatic processing and analysis of articles published by the press. A corpus from The Observatorium database~\cite{7} is studied for evaluating this hypothesis, including articles from several UK newspapers and published online in 2016 and 2017. The questions asked to the Observatorium corpus are the following: What are the concepts and tools needed to describe and explain a political debate? What are narratives on the public opinion space? Specifically, which were the narratives related to Brexit that were propagated in UK along this period of time? 

These issues are here addressed by an integrative scientific approach, incorporating a definition of narrative in social context that borrows theoretical concepts from the social constructivist approach~\cite{8}~\cite{9}, structural linguistics~\cite{10}, and the mathematical theory of Hypernetworks. These high-level concepts are operationalized adapting and evolving state-of-the-art technologies, mainly from the domains of Natural Language Processing and Network Theory, for processing and analysing the Observatorium corpus.

This article introduces the notion of narrative such as described in literature, used as a starting point of a structured, network-based, and operational concept. Afterwards, the context of Brexit is presented through a state-of-the-art of research related to political argumentation. The Observatorium corpus for the case study is described, and the methodology applied to this corpus is proposed and explained in detail. Results depict narratives that were identified within the corpus, its topology and dynamics during 2016 and 2017. Finally, the most relevant results are discussed. 

\subsection{Narratives as communication patterns}

The social constructivist approach in Social Sciences argues that knowledge development results from individuals' interactions~\cite{8}. From this point of view, knowledge is a social construction where narratives, i.e. stories created by humans used for sharing some interpretation of the world, embody meaning and provide common cultural models~\cite{9}. These narratives are shared, and therefore created, transformed, and propagated, using any kind of communication: peer-to-peer, through social networks, or by media broadcast. Different approaches have been proposed to study communication in the social space, such as Qualitative Comparative Analysis, a statistical technique for identifying inferences or implications in data~\cite{12}, System Dynamics, a set of techniques for modelling nonlinear behaviour in systems including flows and feedback loops~\cite{13}, and Social Sequence Analysis, inspired by DNA sequences studied by Bioinformatics and ordering interaction processes~\cite{14}. However, until now narratives in public discourse have hardly been identified and characterized in a quantitative manner, mainly due to the lack of conceptual accuracy concerning their structure and dynamics. 

The concept of narrative that we propose to operate is the result of an integrative science approach, originally influenced by structural linguistics introduced by Saussure, and then developed by Lévi-Strauss and Barthes. The general idea, developed in the mid-20th century, is nowadays attractive again due to the availability of large quantities of data from communication networks. According to Lévi-Strauss, empirical data can be generalized through the understanding of relations between elementary units within discourse, which allow for laws ruling human cultural models to be identified~\cite{10}. Our approach aims at characterizing relations between elementary units in the sense of Lévi-Strauss, together with new developments in Mathematics, namely in Johnsons' Hypernetworks theory, where the basic level relations between elementary components within a network are explained at different abstraction levels, and where aggregated nodes are related at higher levels~\cite{11}. Therefore, the definition of narrative that we propose is the following: 
\begin{displayquote}
\textit{A narrative concerns a social construct where a related and coherent aggregate of terms within public discourse is repeated and propagated on media until it can be identified as a communication pattern, embodying meaning in a way that provides individuals some interpretation of their world.}
\end{displayquote}

This approach, going from discourse elementary units (i.e., words that we will call terms) to its aggregation into topological structures embodying meaning, designed by narratives, and then to dynamics and relations between narratives – is the general idea supporting the methodology proposed in this article. 

However, before going into detailed methodological aspects, an overview on recent research concerning public opinion dynamics is presented. 

\subsection{Uncovering the Brexit public narratives}

Digital spaces, composed by online publications from traditional media, and by social media such as Twitter and Facebook, have been used, since 2016, as a resource for studying the evolution of public opinion and political argumentation around the Brexit debate in UK. Particularly, the Twitter corpus has been used, mainly because of it’s availability to get large quantities of data. The main finding of Twitter-based research concerned the predominance of Euroscepticism on social media during the referendum campaign, where Eurosceptics out-numbered and out-tweeted Remainers. However, Eurosceptics were more confined into their own echo-chambers~\cite{4}. Twitterbots, i.e. Twitter accounts that tweeted during the referendum campaign, disappearing shortly after the ballot, were also considered for characterising retweet cascades and, in general, for typifying propagation phenomena~\cite{15}. The Brexit case study can be compared with others in Europe based on Twitter data, such as the 2017 French presidential election, used to validate an integrated methodology for collecting, reconstructing, analysing, and visualizing the relative positioning of political forces~\cite{16}. Nevertheless, Twitter data is merely representative of the population communicating through this social network, and therefore allows only partial analysis. Facebook data have also been used, for instance for identifying correlations between political interest and online participation on Facebook~\cite{17}. However, the critique regarding restrictions on the Twitter universe can be applied to Facebook as well. Furthermore, the contemporary cacophony from Twitter and Facebook, characterised by getting more of what you like and avoiding exposure to what you disagree with, normalises exceptional political positions that become popular~\cite{31}. These dynamics, as reported by some authors, are based on a new populist architecture of the social media~\cite{5}. 

Other studies, presented either in journalistic or scientific style, are based in press news and opinion articles about Brexit that have been published on newspapers, either in paper or online. However, most of the UK press has supported on side, “Remain” or “Leave”, frequently disseminating a propaganda campaign in which facts and analysis were sacrificed to their ideologically driven goals, propagating distortions and half‐truths~\cite{5}. As a matter of fact, frequently front‐page headlines repeated ideologically charged associations such as ‘EU’ with ‘migrants’, ‘borders’, or ‘refugees’. Traditional content analysis of newspaper data studied the support given to the pro-Leave campaign, or to the Remain arguments, concluding that some newspapers influenced the Referendum, mainly by propagating frames highlighting popular anxieties, such as those regarding migrants~\cite{6}. A classification, proposed by the Loughborough University, organised newspapers not only from their official endorsements, but more importantly based on a qualitative analysis measuring where titles sit on the continuum of opinions between Remain and Leave, and distinguishing between strong or weak support for one of these polarities~\cite{18}. This classification was particularly useful as a starting point for our work.

Two factors explain the limits of the research mentioned in literature. The first one concerns its difficulty to characterize, and then to explain, the evolution of political arguments in public opinion. Namely, a well-defined concept of narrative, as a set of relevant terms and links between these terms, is remarkably missing. Moreover, there is no consensual methodology for the analysis of political arguments dynamics. We argue that a clear concept of narrative, associated with a coherent and state-of-the-art methodology, and a relevant corpus of news and opinion articles, might result in a deeper analysis of the public opinion space. The corpus here being used is as follows.

\section{Data}

\subsection{The Observatorium}

Founded in 2008, the Observatorium research group is dedicated to the collection of textual data from the Internet~\cite{7}. Newspaper articles published on-line are collected in a systematic way. More than one hundred newspapers are currently being gathered, mainly from Europe, but also from other places around the World, from China to Africa, and to South-America. This very large corpus~\footnote{ The Observatorium database was mentioned in 2010 by the MIT Technology Review as “one of the online databases that define our planet”~\cite{19}.} has been used for research, namely in linguistics and economics. Its main goal is to monitor the evolution of knowledge and opinion dynamics in society, namely to understand the structure of arguments and ideas influencing public opinion dynamics.

\subsection{Data for the Brexit case study}

The corpus used in this research is composed by a collection of UK press articles, from selected newspapers, and gathered from the Internet. This collection is complete in the sense that, for this period – 2016 and 2017 - all articles published online, from these newspapers, are included. 

Surveys and analyses have been recently published concerning the political orientations of UK newspapers~\cite{20}~\cite{21}, and how each newspaper have been supporting pro-Brexit or anti-Brexit arguments~\cite{22}. These studies supported the categorisation of the corpus into three political orientations: left-wing (LW), including the Daily Mirror, The Guardian, and The Independent; right-wing (RW), comprising articles from the Daily Star and The Telegraph; and far-right (FR), containing articles from the Daily Express.

All articles related to Europe were selected from the corpus. Selection concerned articles including the term 'Europe' or one of its lemmas: "European Union", "EU", "European Community", "EC", "European Economic Community", "EEC", "Common Market". 

The number of articles under analysis for the period from 2016-01-01 to 2017-12-31, concerning Europe and classified by political orientation, was the following:

\begin{table}[!h]
\centering
\caption{Number of articles under analysis for the period from 2016-01-01 to 2017-12-31}
\label{table_1}
\begin{tabular}{llc}
\hline
{\bf Year} &{\bf Political orientation} &{\bf Articles related to "Europe"} \\
\hline
\multirow{3}{*}{2016} &Left-wing &61246 \\
&Right-wing &11831 \\
&Far-right &13719 \\ 
\hline
\multirow{3}{*}{2017} &Left-wing &57091 \\
&Right-wing &7329 \\
&Far-right &12158 \\ 
\hline
\end{tabular}
\end{table}

The following methodology was applied to the corpus.

\section{Methods}

The goal of the methodology here being proposed is to characterize the topology and dynamics of public discourse narratives. As introduced in Section 1, this is based on the structural linguistics engagement for relating elementary units within discourse~\cite{10}, combined with the general bottom-up mechanics introduced the by hypernetworks theory, where elementary units are aggregated into top level concepts~\cite{11}. The main challenge is how to put together and make operational the high-level concepts from Lévi-Strauss and Johnson. To do it, a diversity of state-of-the-art existing technologies, mainly from the domains of Natural Language Processing and Network Theory, is used for processing and analysing the Observatorium dataset. The basic inputs of the methodology are articles published online by newspapers. High level outputs are narratives. The mechanics allowing to go from a set of unrelated words (bag-of-words) to a narrative featuring some coherent meaning is complex, requiring several incremental steps of observation, analysis, and design. Between the first and the former, methodology steps include topic detection, term inference, depiction of spatial relations between these terms, and finally analysis of its dynamics. A synthesis of these complementary steps and technologies is presented in Table 2. The table can be read top-down, distinguishing tools used for classical topic detection from those depicting narratives, or bottom-up, classifying tools used for illustrating narrative topology vs. narrative dynamics.

\begin{table}[!h]
\centering
\caption{Contributions for an integrated methodology}
\label{table_2}
\begin{tabular}{c|c|c|c|c|c}
  & {\bf Topics} \cellcolor{lightgray} & \multicolumn{4}{c}{ { \bf Narratives } \cellcolor{lightgray!35} }\\
\hline
\makecell{ { \bf Textual and }\\{ \bf  Statistical }\\{ \bf  Algorithm }} 
& \makecell{ Latent \\ Dirichlet \\ Allocation \\ (LDA) } 
& \makecell{ Text Rank \\ + \\ BM25+ } 
& \multicolumn{3}{c}{\makecell{ Term Frequency-Inverse \\ Document Frequency (TF-IDF) \\ + \\ Word2Vec two-layer \\ neural network }} \\
\hline
{ \bf Textual Output } 	
& \makecell{30 topics, \\ inc. 20 keywords \\ per topic} 
& summaries	
& \multicolumn{3}{c}{\makecell{ 50 terms per topic, \\ i.e. 1500 terms \\ representing the corpus } } \\
\hline
\makecell{ { \bf Statistical } \\{ \bf Output } }
& \makecell{keyword co- \\ occurrence} 
& - 
& \multicolumn{3}{c}{\makecell{proximity \\ between terms}} \\
\hline
\makecell{ { \bf Graphical }\\{ \bf  Algorithm } }
& \makecell{Force directed \\ (Verlet)}
& \makecell{-}
& \makecell{Force-\\directed \\ (Fruchterman-\\Reingold)}
& \makecell{Sankey}
& \makecell{Linear \\ regression \\ with \\ marginal \\ distribution} \\
\hline
\makecell{ { \bf Graphical }\\{ \bf  Output } }
& \makecell{Topic \\ Network \\ Graph}
& \makecell{-}
& \makecell{Narrative \\ Network \\ Graph}
& \makecell{Narrative \\ Flow \\ Diagram}
& \makecell{Scatter Plots \\ + Histograms } \\
\hline
&  \multicolumn{3}{c|}{ { \bf Topology } \cellcolor{lightgray!35} }
& \multicolumn{2}{c}{{ \bf Dynamics } \cellcolor{lightgray} } \\
\end{tabular}
\end{table}

Original contributions of this methodology concern the sequence of existing state-of-the-art algorithms, where the output of each one becomes the input of the following one, until achieving the design the we propose for Topic Network Graphs, Narrative Network Graphs and Narrative Flow Diagrams (see Table 2). The different steps of the methodology are detailed in the following.

\subsection{Narrative topology}

Given the corpus, the first step regards topic detection through the classical unsupervised algorithm Latent Dirichlet Allocation (LDA)~\cite{23}, where topics are deduced from clusters of words that co-occur in documents. LDA provides a mix of topics composing each document. The advantage of this algorithm is its efficiency when applied to a large corpus. Two results are aimed: identifying which topics are relevant, and quantifying the importance of each topic. These results allow for a first graphical representation of co-occurrence relations between topics, depicted in a Topic Network Graph. This graph is designed using a force-directed graph algorithm that implements a velocity Verlet numerical integrator for simulating physical forces linking nodes~\cite{24}. 

Given a set of topics, next step characterizes its content, i.e. a meaningful sequence of terms in each topic, that we will call narrative. Two procedures are tested – summaries and keyword terms extraction. In the first one, each summary is extracted from a selection of the ten most relevant texts within a topic. A variation of the TextRank algorithm for unsupervised automatic summarization of texts is used, where TextRank is combined with the BM25+ information retrieval algorithm~\cite{25}. The resulting summaries are the expression of narratives in a textual form. However, this description of narratives is only partial, since it doesn't reveal their structure. Namely, summaries don’t allow quantifying the evolution of narratives between two periods of time. 

Considering the limited explanatory output of classical algorithms such as LDA and TextRank, we looked for a largest set of keyword terms for explaining narrative structure and dynamics. The TF-IDF - Term Frequency-Inverse Document Frequency algorithm was used for extracting these terms~\cite{26}. TF-IDF scores the weighs of a term’s frequency (TF) and its inverse document frequency (IDF), and then computes the product of these two scores. The higher this product, the rarer the term, and vice versa. This way terms are assigned the importance they have in the corpus. Our adaptation of the original TF-IDF algorithm~\cite{26} is computed as following. For a term \textit{t} in a newspaper article \textit{a}, the weight \textit{W$_{{k,a}}$} of term \textit{t} in article \textit{a} is given by 
\newline
\[ W_{t,a} = TF_{t,a} log(\frac{N}{DF_{t}}) \]
\newline
where \textit{TF$_{{k,a}}$} is the number of occurrences of \textit{t} in article \textit{a}, \textit{DF$_{{k}}$} is the number of articles containing the term \textit{t}, and \textit{N} is the total number of newspaper articles in the corpus. Fifty terms per each of the 10 topics, i.e. 500 terms for the whole corpus, were extracted from the corpus using the TF-IDF algorithm. In practice, from the previous topic detection step, articles within the corpus are already ordered by relevance according to each topic. We take the \textit{n} most relevant articles for each topic. These most relevant articles per topic are then used as input for computing fifty TF-IDF terms representing each topic. As result, each narrative is represented by fifty terms. This approach provides a broader description than the previous classical LDA Keywords. Relations between terms are then used as input for the next step, depicting narrative’s structure. 

The graphical illustration of narrative’s structure is based on a measure of proximity between its terms. According to an updated state-of-the-art in the domain, a complete and performative way of calculating proximity between terms within a corpus is through the Word2Vec algorithm ~\cite{27}. This is a two-layer neural network trained to identify the context of terms. Generically, Word2Vec maps each word of a corpus in a vector of words according to its context within the corpus. Then, similarity between context of different terms can be evaluated. The result is expressed in a table including the measure of similarity for each pair of terms. Finally, pairs of terms with similar context more than a certain threshold are linked in a Narrative Network Graph. The Fruchterman-Reingold force-directed graphical algorithm~\cite{29} is then used to place nodes in this circular graph. Nodes with week context similarity are placed in the exterior of the circular graph, and vice-versa, i.e. nodes with strong context similarity are placed in the center of the graph. This design allows the distinction between central and marginal terms within a narrative.  

\subsection{Narrative dynamics}

Besides topology, narratives can be considered as a flow of concepts through articles during a certain time period. This methodology keeps the trace of terms from the beginning to the end of the period, and from one narrative to another. Flows of terms are represented in Narrative Flow Diagrams using Sankey diagrams~\cite{30}, in which the width of the arrows is shown proportionally to the term occurrence. Finally, scatter plots and histograms are combined for comparing term frequencies in 2016 and 2017. Calculating, fitting, and plotting the linear regression with marginal distributions concerning the frequency of the terms complete the analysis of the evolution of narratives.

\section{Results}

The methodology above was applied to the Observatorium corpus, including more than 160.000 UK newspaper articles. These data allowed analysing the topology and dynamics of narratives related to “Europe”, been propagated during 2016 and 2017 on the newspapers Daily Mirror, The Guardian, and The Independent (left-wing), Daily Star and The Telegraph (right-wing), and Daily Express (far-right).

\subsection{Narrative topology}

Firstly, topological analysis identified topics through the LDA - Latent Dirichlet Alocation algorithm. Topic keywords were used for designing a Topic Network Graph for each political trend.

\subsubsection{Topic network graphs}

The LDA keywords defining each topic were identified. Articles from each political trend were represented by ten topics, by order of relevance, i.e. by order of occurrence in the corpus. For instance, the first, and therefore most relevant, far-right topic is defined by the following keywords: 

\begin{table}[!h]
\centering
\caption{Example of LDA topic keywords}
\label{table_3}
\begin{tabular}{cc}
\hline
{\bf Topic number} &{\bf Topic keywords} \\
\hline
1 & \makecell{ 
    migrants migrant europe germany merkel people turkey german country \\ 
    year refugees crisis border related asylum countries refugee italy france
} \\
\hline
\end{tabular}
\end{table}

Annex 7.1 lists all LDA topic keywords for far-right, right-wing, and left-wing.

Relationships between different topic keywords were then depicted within Topic Network Graphs, where each node symbolises a keyword, and colours of nodes indicate topics. Hot colours, such as red, orange, or yellow, are given to the most frequent topics within the corpus, while cold colours, such as green or blue, indicate less frequent topics. Also, two nodes are connected if their co-occurrence statistics within the corpus is more than a given threshold. Figures 1, 2, and 3 exemplify Topic Network Graphs for the different political tendencies during 2016. The complete set of images for these graphs, including 2016 and 2017, is available in Annex 7.2.

\begin{figure}[!h]
\centering
\includegraphics[width=3.5in]{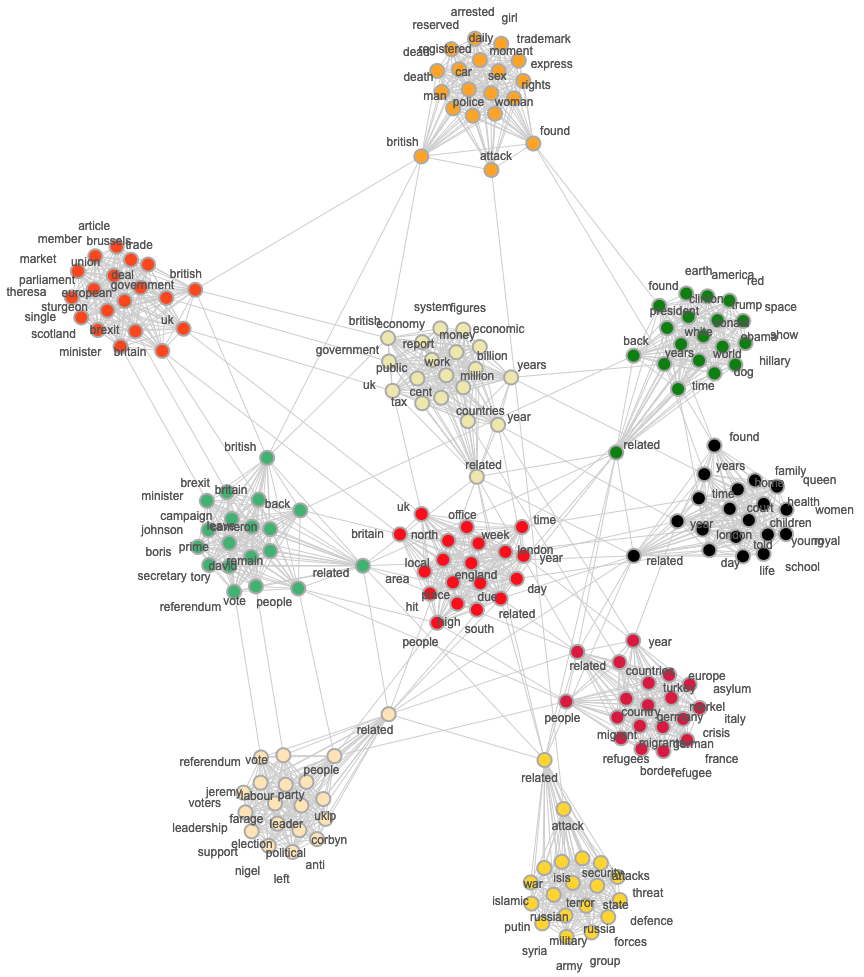}
\caption{Example of a Topic Network Graph: far-right, 2016}
\label{fig:1}
\end{figure}

The main subjects related to Europe are patent in the Topic Network Graphs as follows. Far-right (Figure 1) is focused on relating emigration to specific countries, such as Germany, Turkey, Italy, and France. For this political wing, the relation between emigration and refugees is also present. During 2016, far-right related Europe to attacks against women and children, and sexual offenses. The notion of Europe was also associated to ISIS terrorism. Economics and internal policy appeared in less important topics; however, its relevance tended to raise during 2017. 

\begin{figure}[!h]
\centering
\includegraphics[width=4.5in]{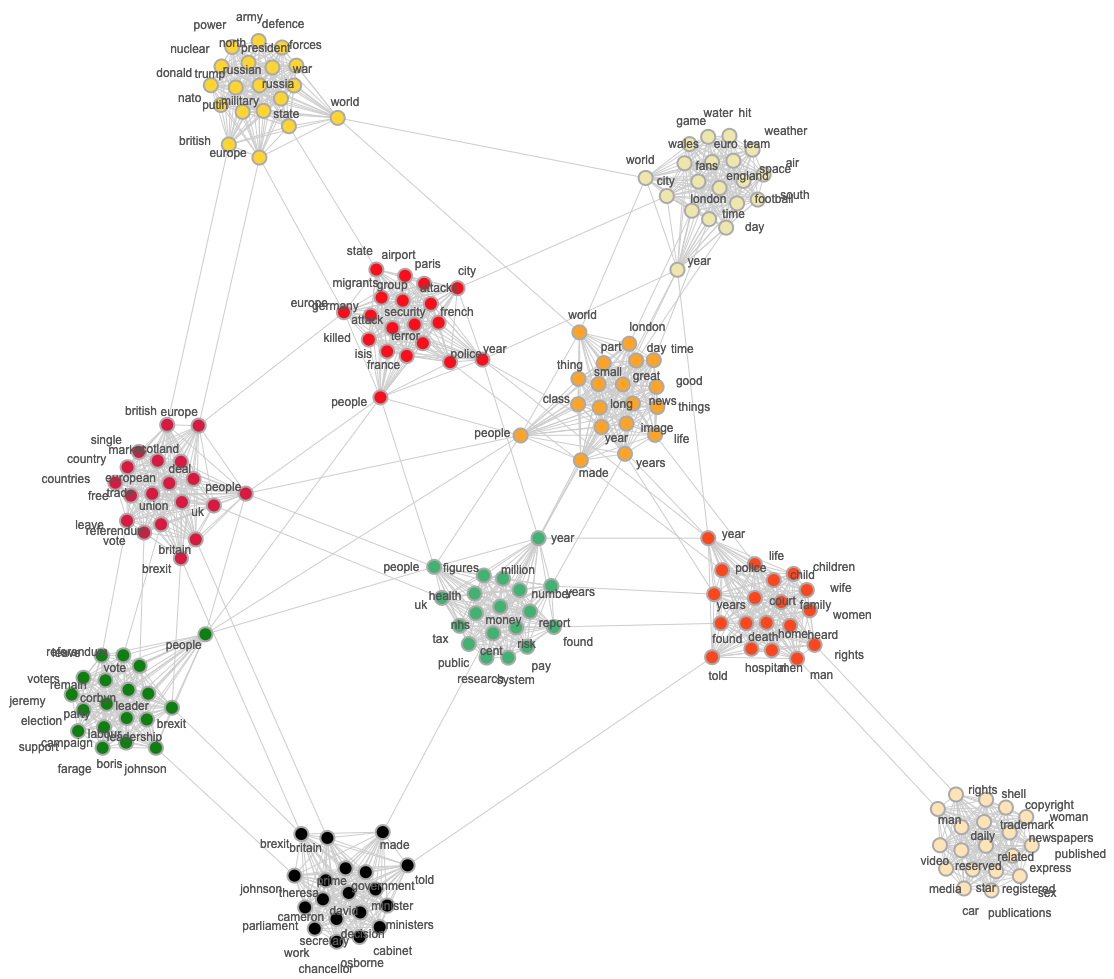}
\caption{Example of a Topic Network Graph: right-wing, 2016}
\label{fig:2}
\end{figure}

Concerning right-wing, depicted in Figure 2, its focus was mainly internal politics (brexit, government, minister, parliament, scotland) and, less importantly, external politics (trump, president, russia, korea, putin, security, world, defence, white house).

\begin{figure}[!h]
\centering
\includegraphics[width=3.5in]{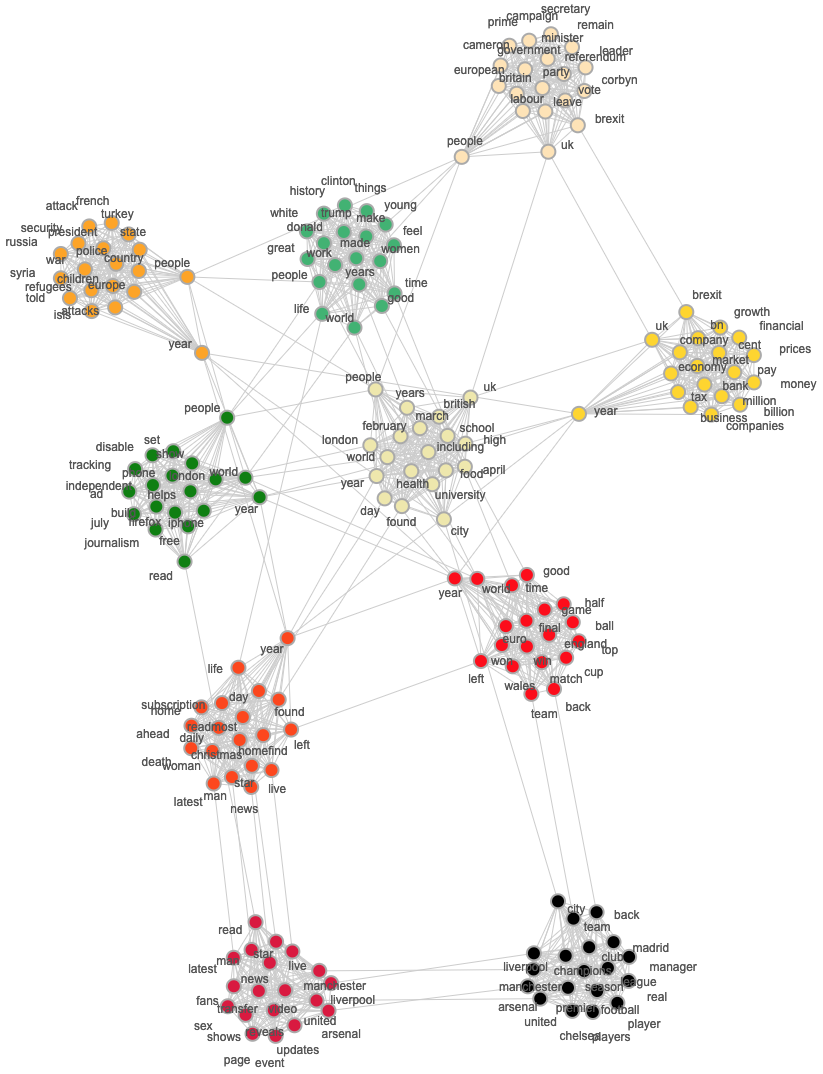}
\caption{Example of a Topic Network Graph: left-wing, 2016}
\label{fig:3}
\end{figure}

 Regarding left-wing, presented in Figure 3, surprisingly the most relevant topics were related to football (live, united, manchester, liverpool, arsenal, fans) and to people’s news (christmas, day, life, home, woman, children, health, social). In 2016, refugees, ISIS, terrorism and international politics appeared on the 4th place. Economic discussion emerged only on the 6\textsuperscript{th} place. During 2017 the trend was generally similar, however terrorism dropped to a less relevant 10\textsuperscript{th} place (see Annexes 7.1 and 7.2 for the complete collection of topic keywords and graphs).

\subsubsection{Summaries}

The main weakness of topic detection is the simplicity of the analysis it provides. The most recent advances in corpus analysis includes summarisation techniques, such as the TextRank algorithm for unsupervised automatic summarisation of texts. This technique was used for extracting 30 summaries, one for each topic on the different political trends. As an example, the summary of the first far-right topic in 2016 is the following:

\begin{displayquote}
\textit{ The news comes after sex attacks by gangs of 'migrants' in Cologne on New Year's Eve AFP117 women were sexually assaulted and robbed in Cologne on New Year's EveIn the wake of the Cologne attacks, the German Chancellor promised to give authorities more powers to crack down on migrants who commit crimes, including deporting them. She admitted Europe was "vulnerable" in the refugee crisis because it was not yet in control of the situation to the extent that it would like to be. Speaking at a business event near Frankfurt, she said: "Now all of a sudden we are facing the challenge that refugees are coming to Europe and we are vulnerable, as we see, because we do not yet have the order, the control, that we would like to have."As Europe has struggled to cope with the refugee crisis, some European Union (EU) countries have re-established border controls within the passport-free Schengen zone, where they had been abolished. Mrs Merkel said that to preserve the Schengen zone within the EU it was necessary to make the bloc's external borders more secure. A police spokeswoman in Munich confirmed Germany sent back up to 1 but as the government takes a tougher stance, German interior minister Thomas de Maizier warned: "I expect that in the next weeks, the number of repatriations, voluntary returns and deportations will rise significantly." Related articles Now BRITAIN faces migrant SEX ATTACKS after surge of European assaults, warns top expert. }
\end{displayquote}

The complete collection of summaries is available in Annex 7.3. Summaries interestingly express narratives in its textual form. However, this doesn't reveal the conceptual structure of narratives. Namely, their evolution between two moments in time isn't characterised. Therefore, topology and dynamics will require going deeper into the research, as follows. 

\subsubsection{Terms}

According with the methodology previously explained, a large set of terms representing each narrative was synthesised using TF-IDF in its adapted version to the Observatorium corpus. This algorithm allows exploring the relations between terms, and therefore the structure of each narrative. Results include fifty terms for each topic and a total of 10 topics * 50 terms = 500 terms for each political tendency, i.e. a total of 1500 terms for the whole corpus. The complete set of terms is listed in Annex 7.4.  

\subsubsection{Topology}

The topological structure of narratives is evinced in Narrative Network Graphs through the combination of two algorithms, the two-layer neural network Word2Vec and the Fruchterman-Reingold force-directed graphical algorithm (both previously mentioned in Section 3). The advantage of such representation is that it allows distinguishing terms sharing a common context, that are placed in the centre of the graph, from those with a specific context, positioned in its periphery.

\begin{figure}
\centering
\begin{subfigure}{.5\textwidth}
  \centering
  \includegraphics[width=1.0\linewidth]{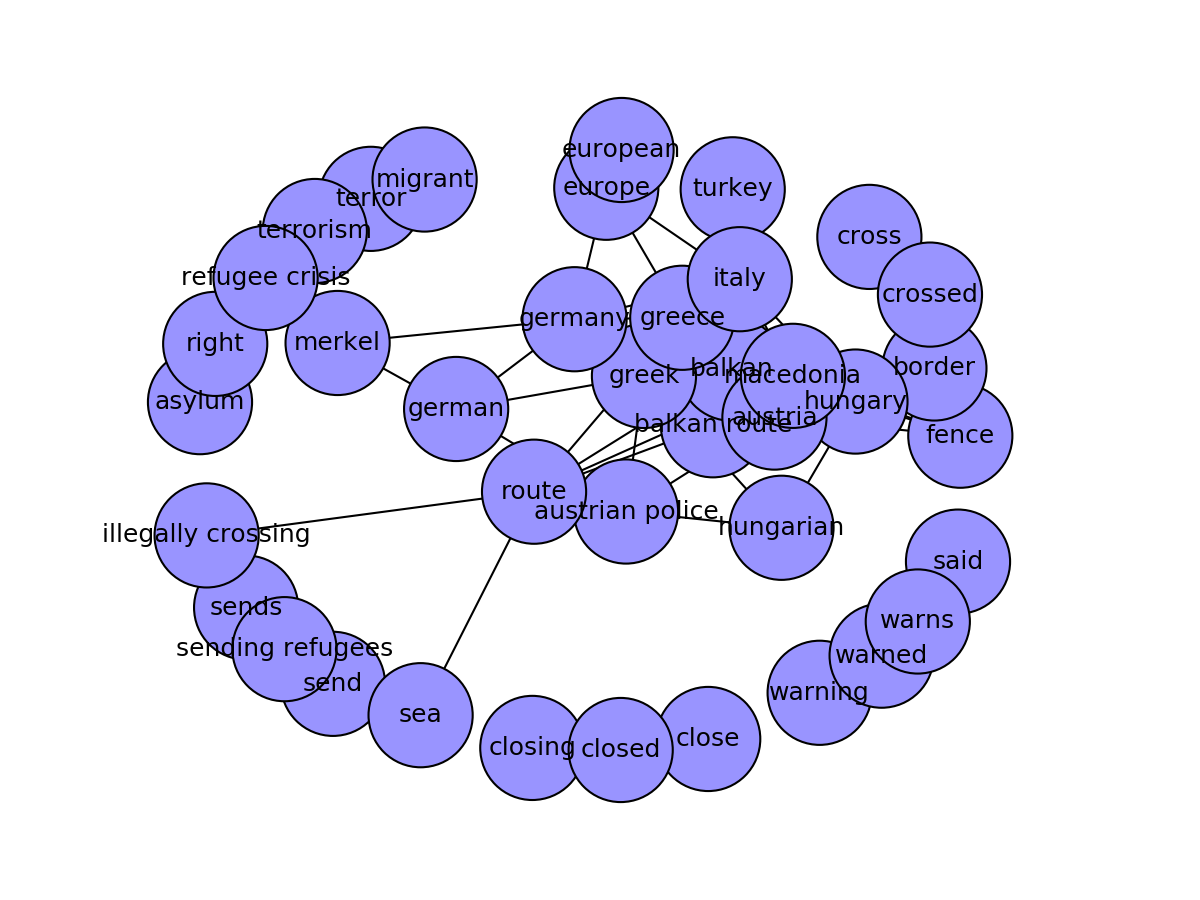}
  \caption{1 - migrants \textunderscore  refugees}
  \label{fig:4sub1}
\end{subfigure}%
\begin{subfigure}{.5\textwidth}
  \centering
  \includegraphics[width=1.0\linewidth]{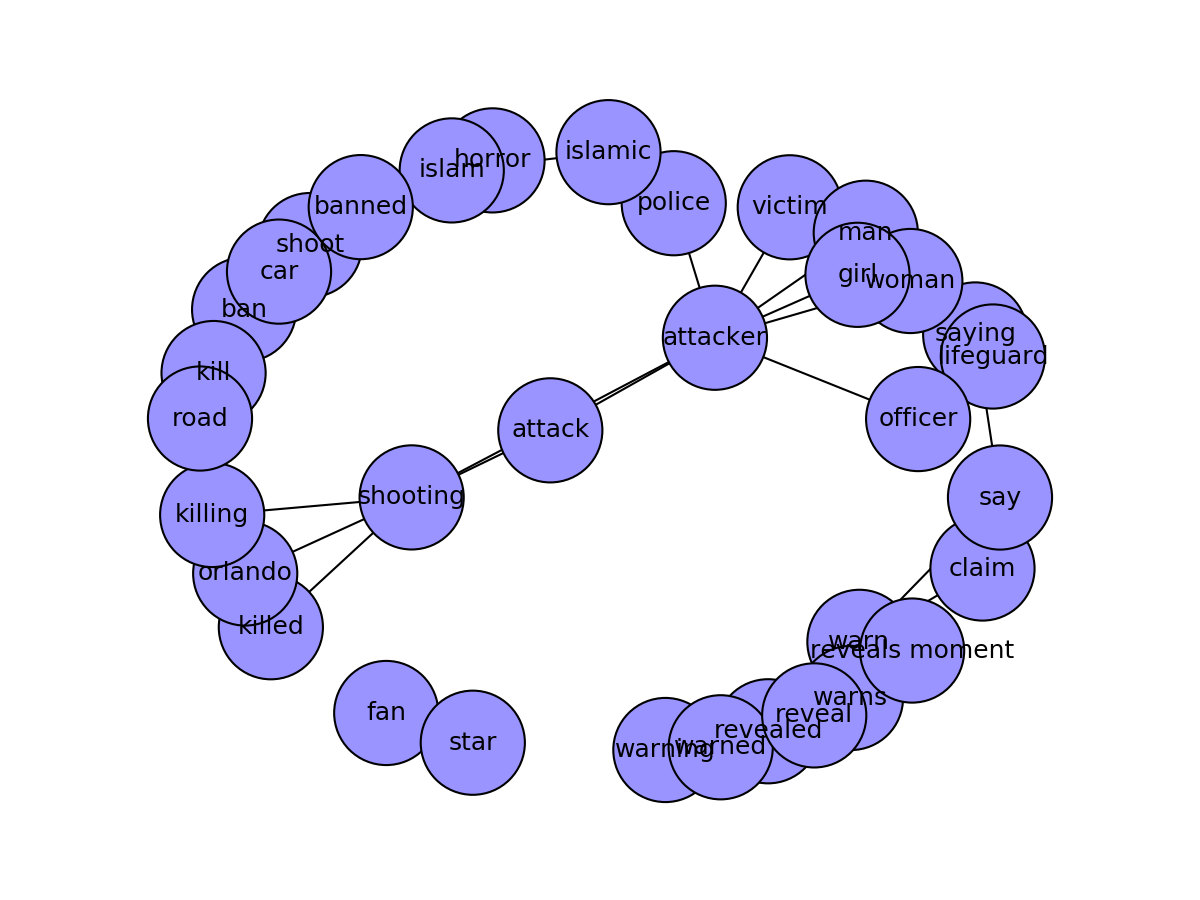}
  \caption{4 - attacks\textunderscore  woman\textunderscore  migrants}
  \label{fig:4sub2}
\end{subfigure}
\caption{Narrative Network Graphs: examples of two far-right narratives in 2016}
\label{fig:4}
\end{figure}

Figure 4 exemplifies two far-right narratives during 2016. The first one, entitled “1 – migrants\textunderscore refugees”, represents concepts related to the path of emigrants crossing European countries such as Austria, Greece, or Hungary, as well as an emigration destination country, Germany. All these terms have common contexts within the emigration debate, and therefore they are positioned in the center of the graph. Otherwise, terms such as “terror” and “asylum”, on one side, and “cross”, “border”, and “fence”, on the other side, have specific contexts, and therefore are peripheral. The second example, named “4 – attacks\textunderscore woman\textunderscore migrants”, represents central terms “shooting”, “attack” and “attacker”, and a diversity of peripheral terms composing two sub-narratives, such as “islam” and “terror” for the one, and “victim”, “man”, and “woman” for the other. Sub-narratives are visually identified by groups of close, interconnected terms.

\begin{figure}
\begin{subfigure}{.5\textwidth}
  \centering
  \includegraphics[width=1.0\linewidth]{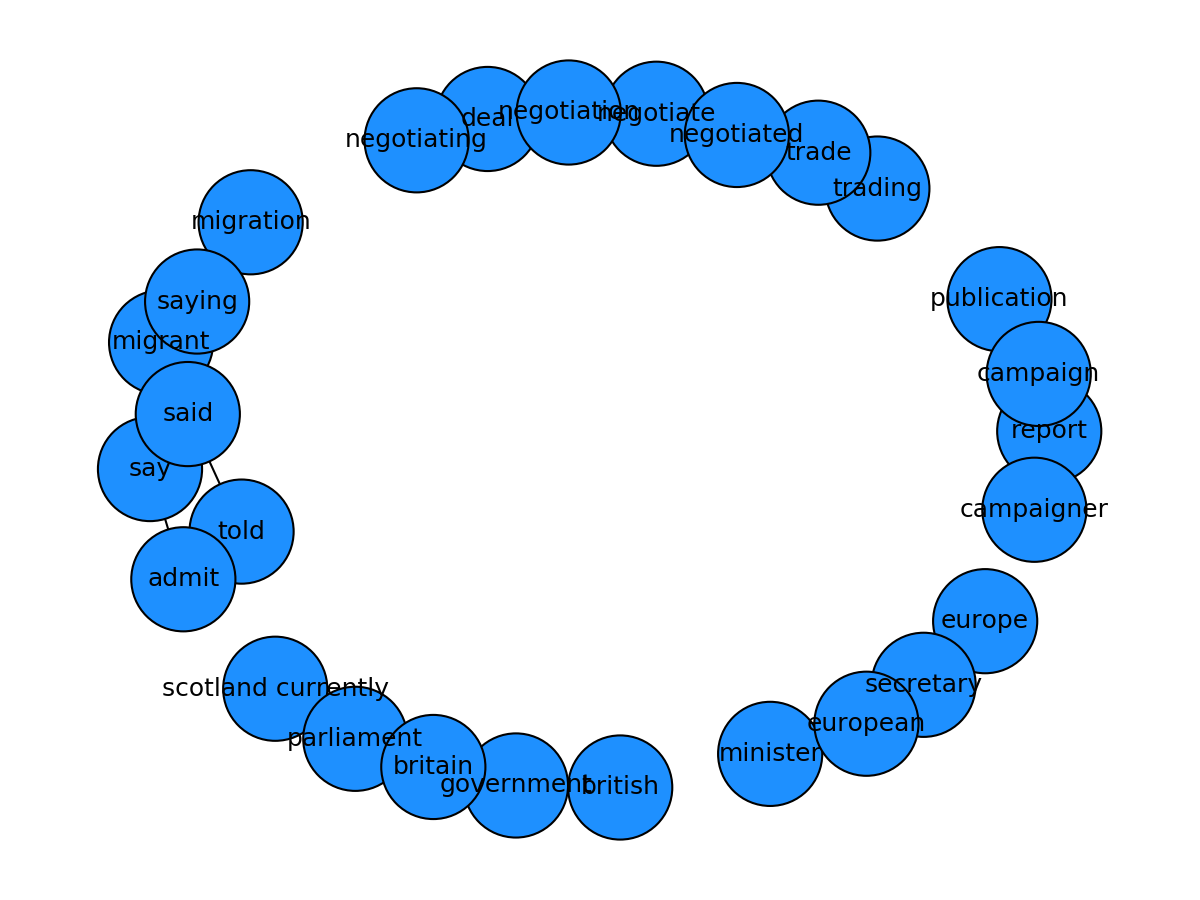}
  \caption{1 - referendum\textunderscore britain\textunderscore scotland}
  \label{fig:5sub1}
\end{subfigure}%
\begin{subfigure}{.5\textwidth}
  \centering
  \includegraphics[width=1.0\linewidth]{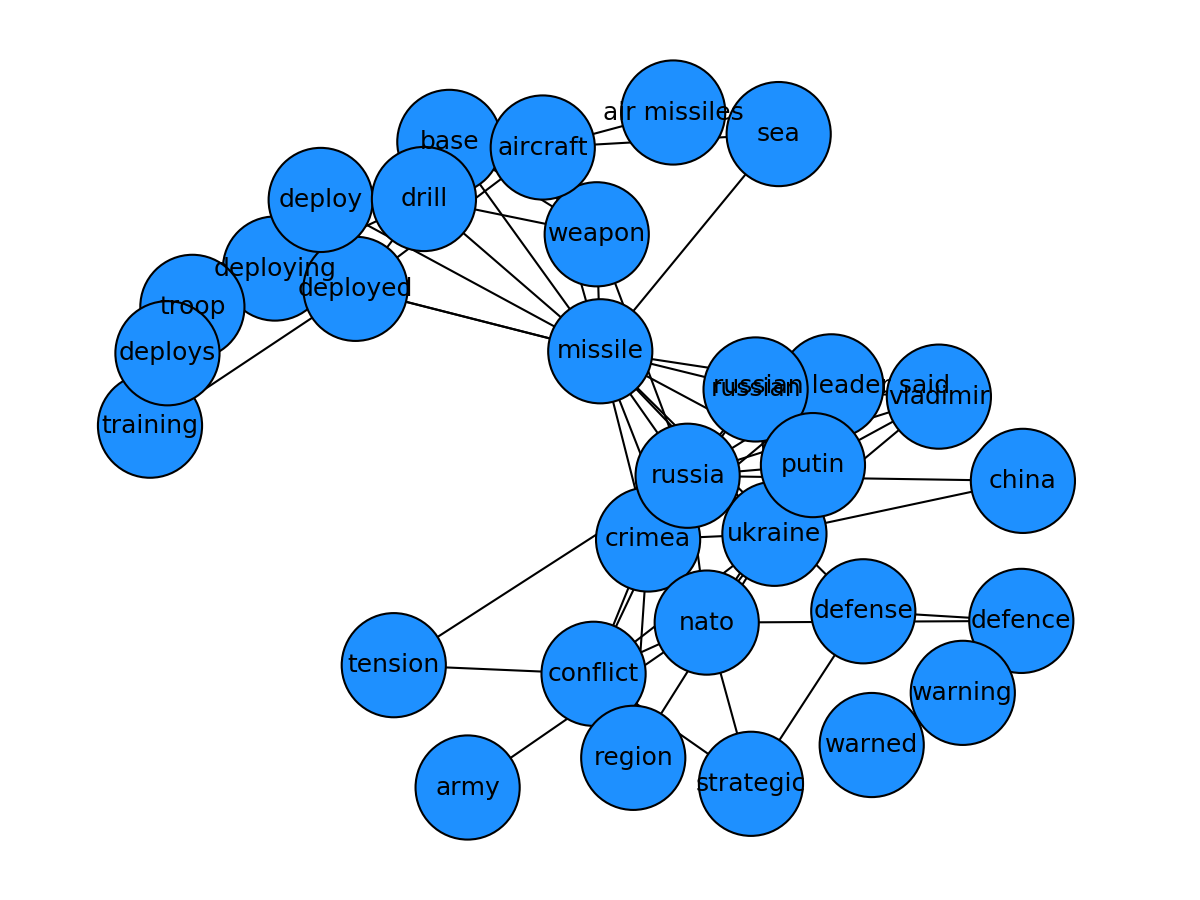}
  \caption{6 - trump\textunderscore russia\textunderscore military}
  \label{fig:5sub2}
\end{subfigure}
\caption{Narrative Network Graphs : examples of two right-wing narratives in 2016}
\label{fig:5}
\end{figure}

Figure 5 reveals two right-wing narratives. The first one, called “1 – referendum\textunderscore britain\textunderscore scotland”, includes only peripheric terms. These are grouped into several sub-narratives, such as one related to “negociation”, and another one to “campaign”. The second narrative “6 – trump\textunderscore russia\textunderscore military” has a specific topology, where the central term “missile” is the link between a sub-narrative related to military technology, including the terms “weapon”, “aircraft”, “deploy”, and “base”, and another sub-narrative includes terms representing international actors and countries – “putin”, “russia”, “ukraine”, “china”, etc. 
\begin{figure}
\begin{subfigure}{.5\textwidth}
  \centering
  \includegraphics[width=1.0\linewidth]{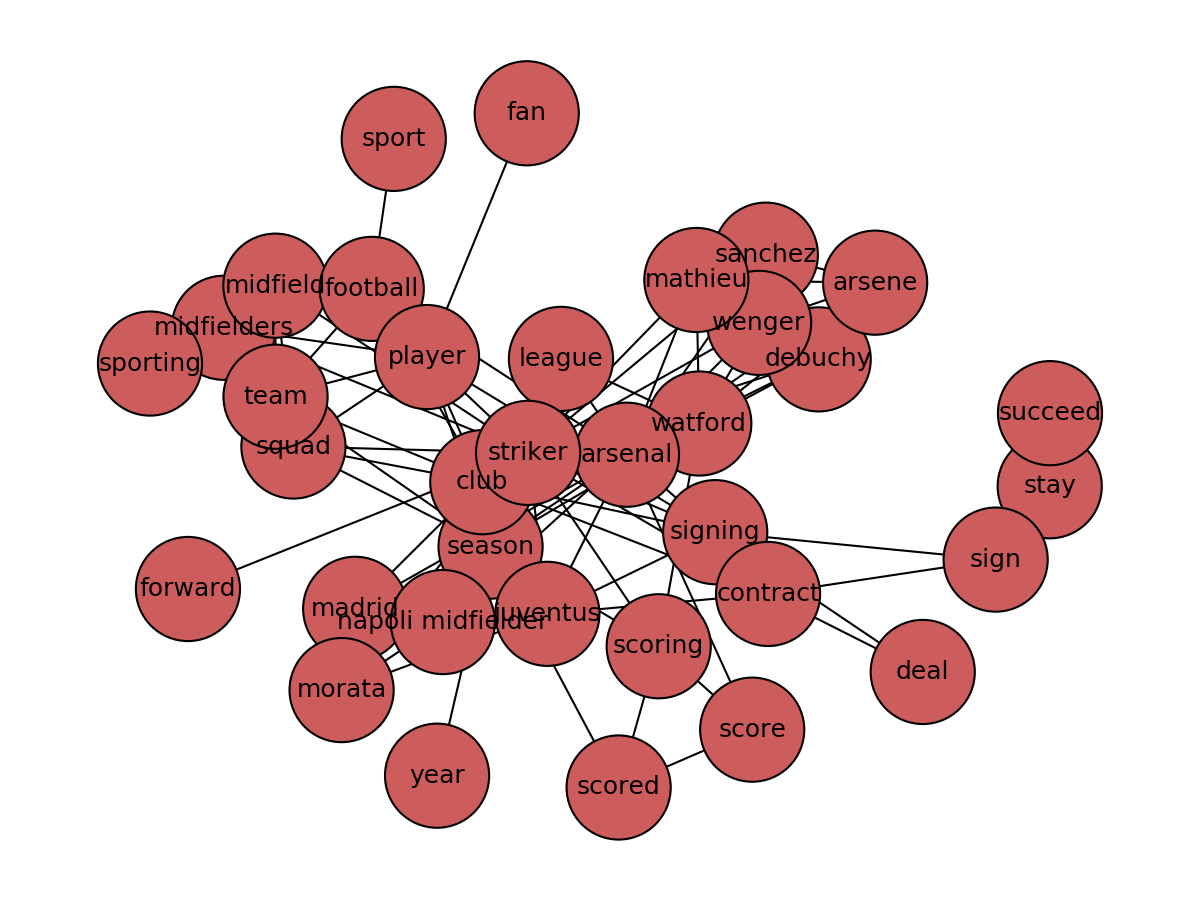}
  \caption{5 - football}
  \label{fig:6sub1}
\end{subfigure}%
\begin{subfigure}{.5\textwidth}
  \centering
  \includegraphics[width=1.0\linewidth]{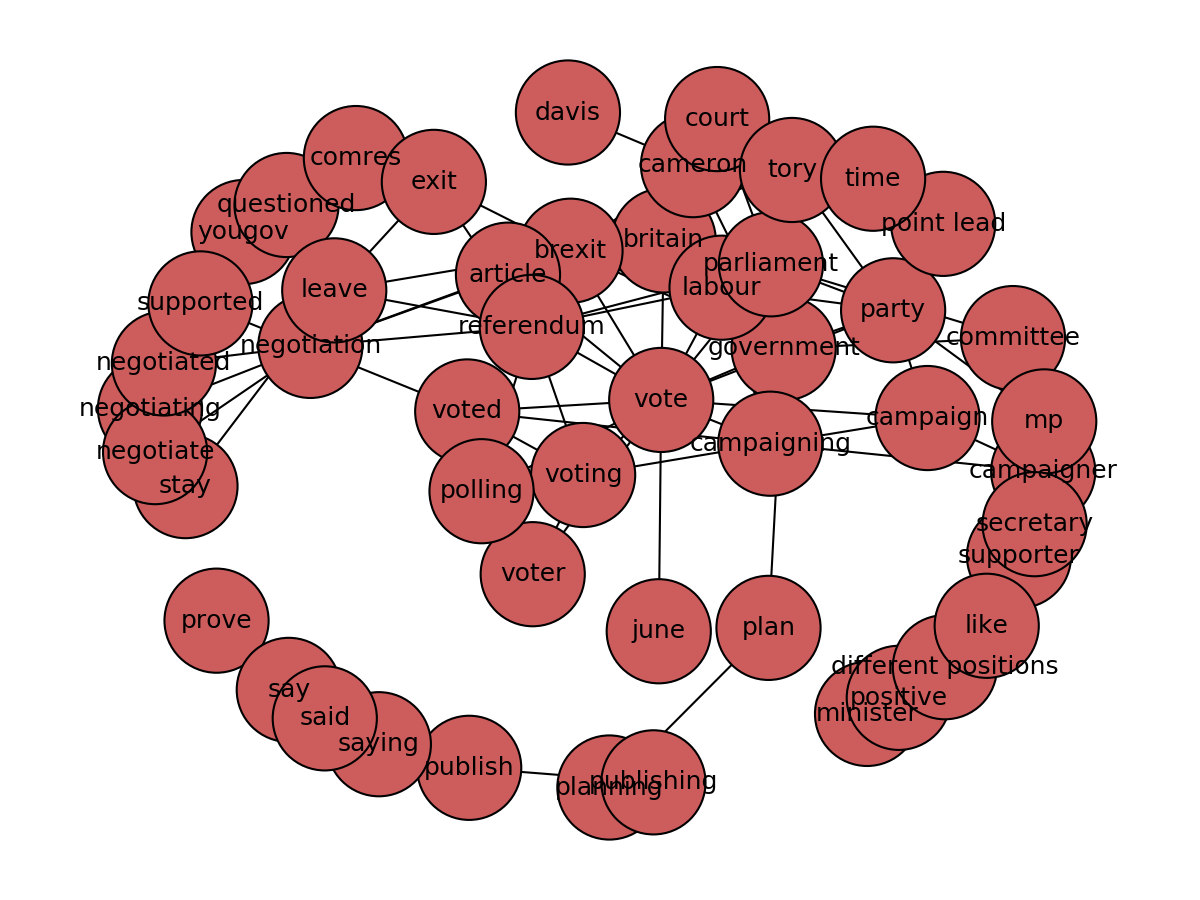}
  \caption{8 – brexit\textunderscore referendum}
  \label{fig:6sub2}
\end{subfigure}
\caption{Narrative Network Graphs: examples of two left-wing narratives in 2016}
\label{fig:6}
\end{figure}

Finally, Figure 6 depicts two examples of left-wing narratives. The first one, with the title “5 – football”, presents a mixed topology, with several central terms and crossed links. Next example, “8 – brexit\textunderscore referendum”, has a clearer nucleous (“vote”, “referendum”, “campaign”) and peripheral terms groupe into several sub-narratives. 

The complete set of Narrative Network Graphs is available in Annex 7.5. 

\subsection{Narrative dynamics}

\subsubsection{Flows of narratives}

Also according with the methodology previously described, analysing the dynamics of narratives implies characterising the flow of terms during a given period. This means observing how a given narrative evolves, if it grows, or keeps its relevance, or if, on the contrary, it is diluted into less relevant ones. This analysis is done using Sankey diagrams for designing Narrative Flow Diagrams, in which the width of the arrows is shown proportionally to the flow quantity. 

\begin{figure}[!h]
\centering
\includegraphics[width=5.0in]{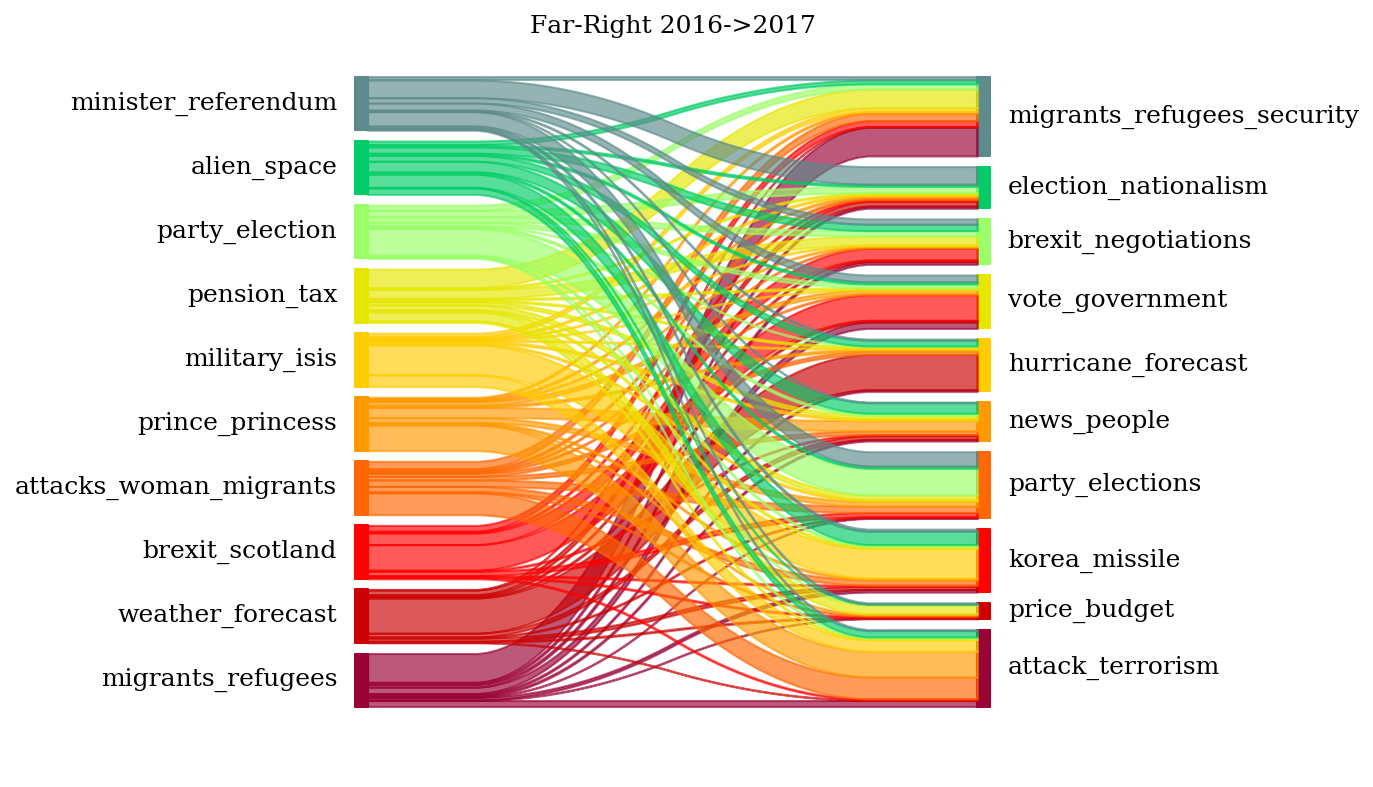}
\caption{Narrative Flow Diagram: Far-right, from 2016 to 2017}
\label{fig:7}
\end{figure}

Three examples related to the three political tendencies are presented. The first one concerns far-right. Figure 7 depicts narratives as it were characterised in 2016 (on the left of the graph) and in 2017 (on the right of the graph). Most relevant narratives, i.e. those most frequent in the corpus, are represented using hot colours (red, orange, yellow), from the bottom to the top of the graph. The most relevant one, in 2016, is named “migrant\textunderscore refugees”, the next most relevant one is “weather\textunderscore forecast”, and so on. From 2016 to 2017 terms might move from one narrative to another one, or more. For example, a certain number of terms from the 2016 “migrant\textunderscore refugees” narrative appear in the last 2017 “migrants\textunderscore refugees\textunderscore security” narrative. This means that, for far-right, the importance of the narrative related to migrants and refugees was important during the 2016 referendum campaign, but has dropped considerably during 2017. Otherwise, a new narrative appeared in 2017, in first place, regarding “attack\textunderscore terrorism”. Particularly noteworthy is that this new one is partially composed by the 2016 “attacks\textunderscore woman\textunderscore migrants” terms.

\begin{figure}[!h]
\centering
\includegraphics[width=5.0in]{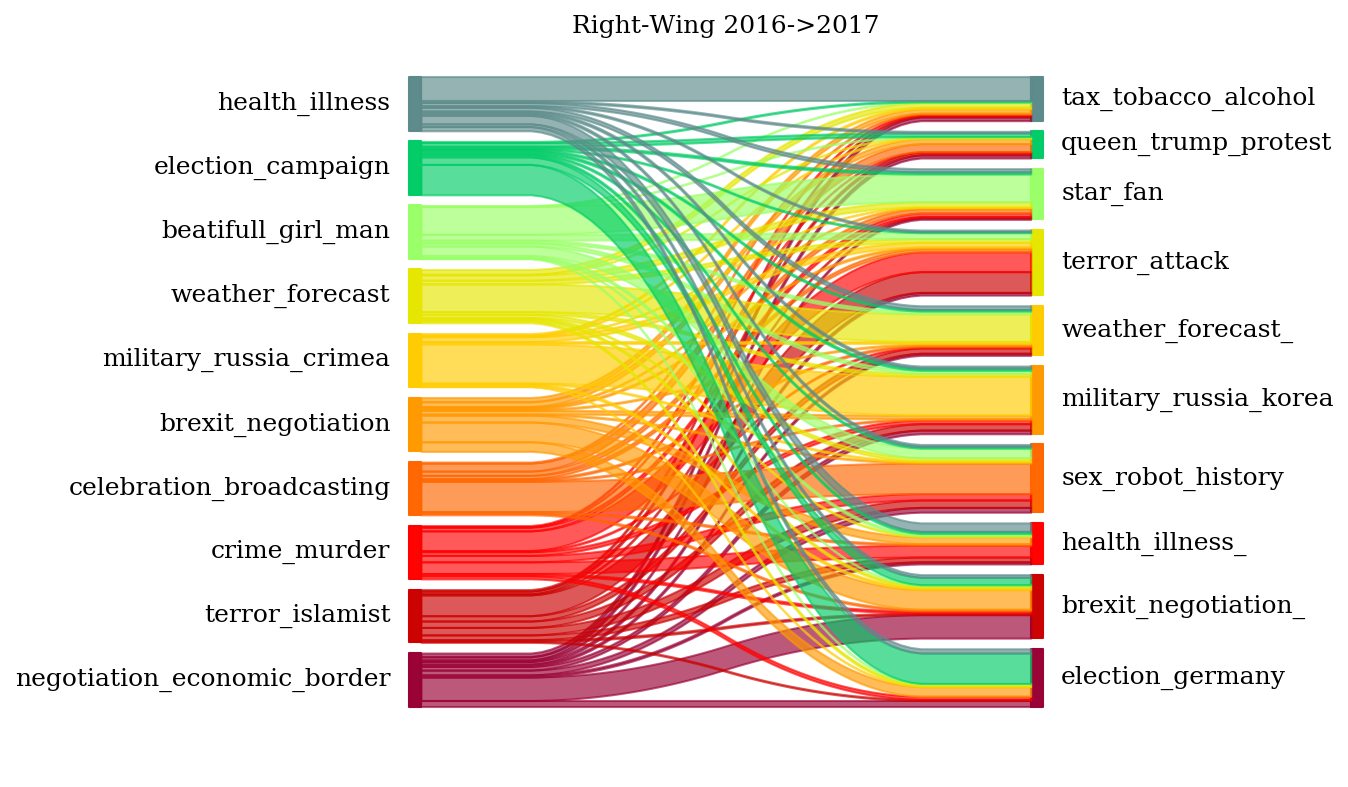}
\caption{Narrative Flow Diagram: Right-right, from 2016 to 2017}
\label{fig:8}
\end{figure}

The right-wing flow of narratives (Figure 8) shows stability around “negociation\textunderscore economic\textunderscore border” and “brexit\textunderscore negotiation” (first and second positions bottom-up on the graph). “election\textunderscore germany” is a new 2017 narrative (first place), and 2016 “health\textunderscore illness” raised importance in the period, attaining position 3 in 2017. Otherwise, the 2016 second place “terror\textunderscore islamist” evolved to a 2017 seventh place “terror\textunderscore attack”, which means a decrease of relevance for this narrative in right-wing data.

\begin{figure}[!h]
\centering
\includegraphics[width=5.0in]{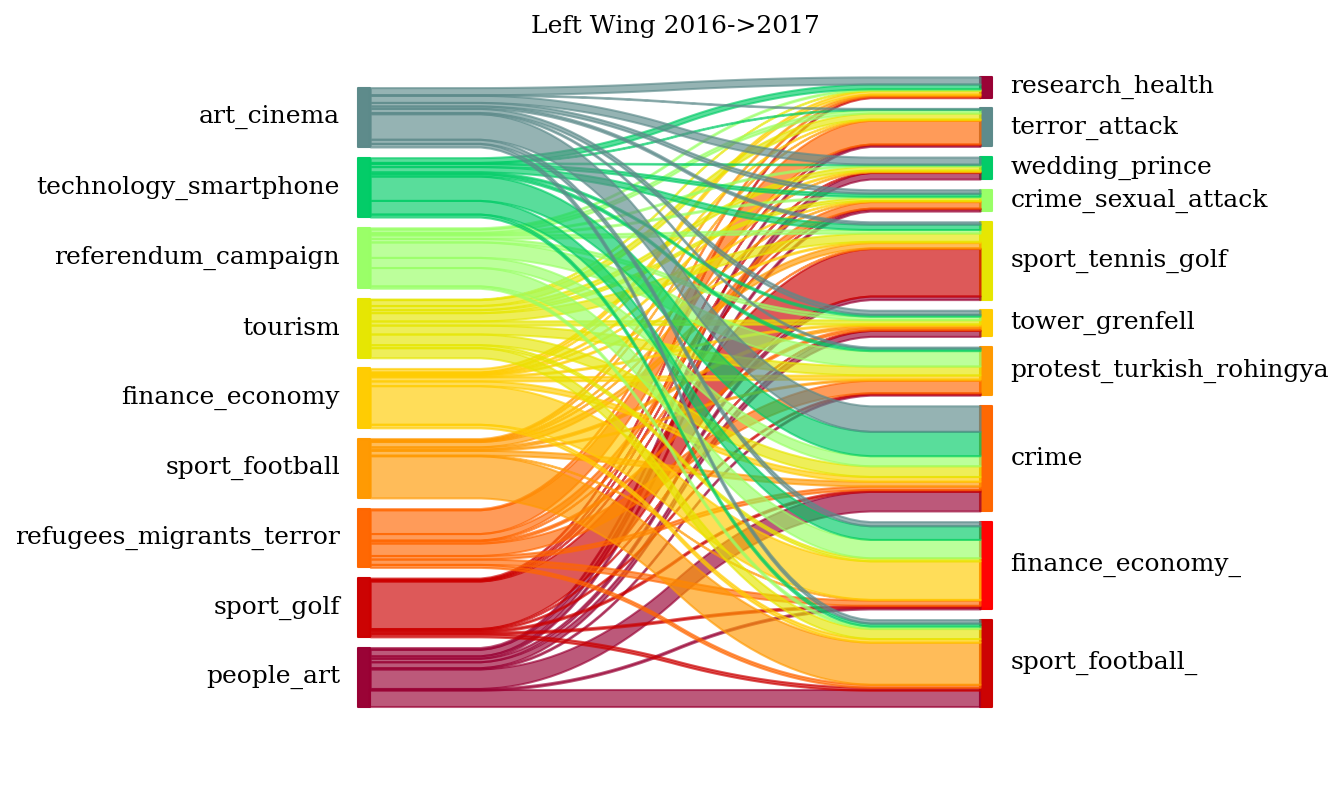}
\caption{Narrative Flow Diagram: Left-right, from 2016 to 2017}
\label{fig:9}
\end{figure}

Finally, left-wing flows (Figure 9) express a different kind of dynamics. During 2016 the most important ones, related to “Europe”, were non-political, such as “people\textunderscore art” (first place bottom-up) and “sport\textunderscore golf” (second place bottom-up). The first political narrative, “refugees\textunderscore migrants\textunderscore terror”, appears on the third place. However, during 2017 subjects related to “crime” and “finance\textunderscore economy” raised importance. 

\subsubsection{Evolution of term usage}

Finally, the analysis of the evolution of narratives is completed by calculating, fitting, and plotting scatter plots and histograms, in here combined to compare frequencies of terms in 2016 and in 2017. 

\begin{figure}
\centering
\begin{subfigure}{.333\textwidth}
  \centering
  \includegraphics[width=0.9\linewidth]{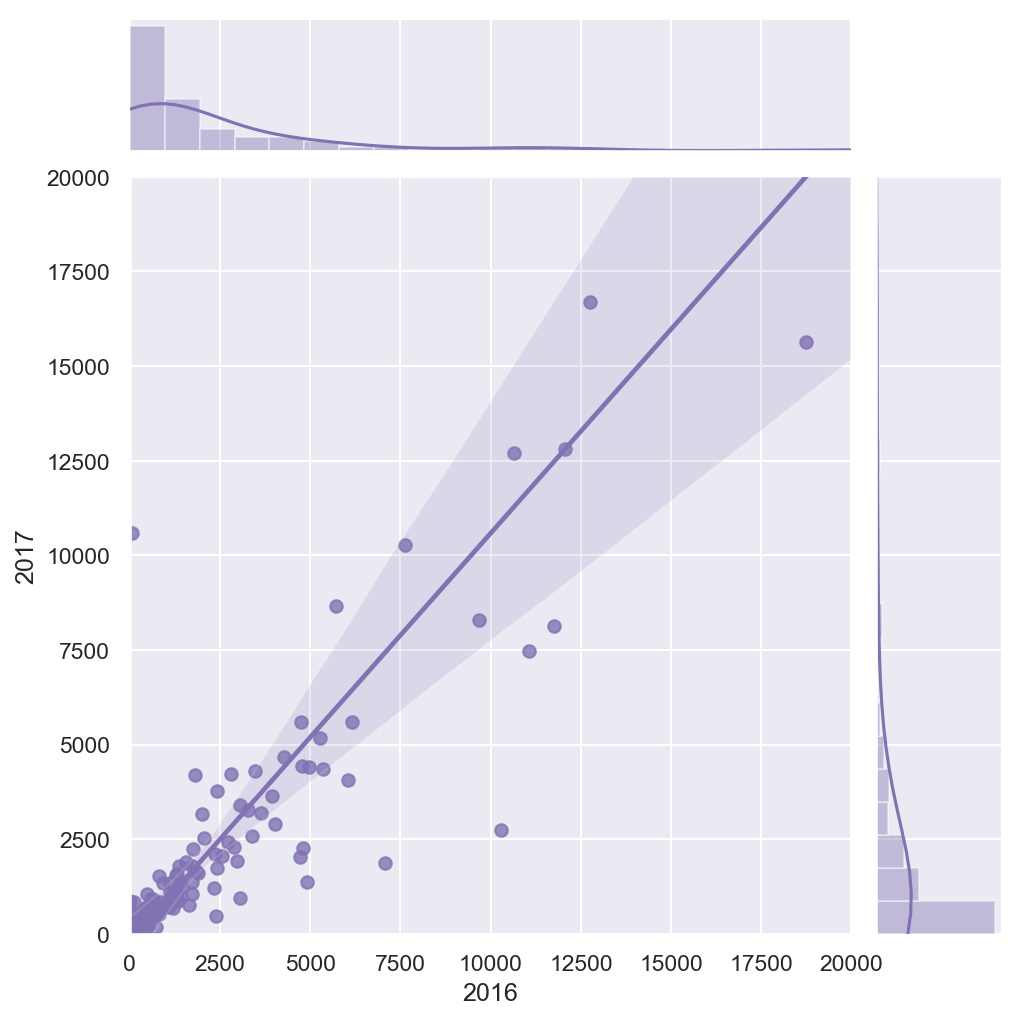}
  \caption{Far-right}
  \label{fig:10sub1}
\end{subfigure}%
\begin{subfigure}{.333\textwidth}
  \centering
  \includegraphics[width=1.0\linewidth]{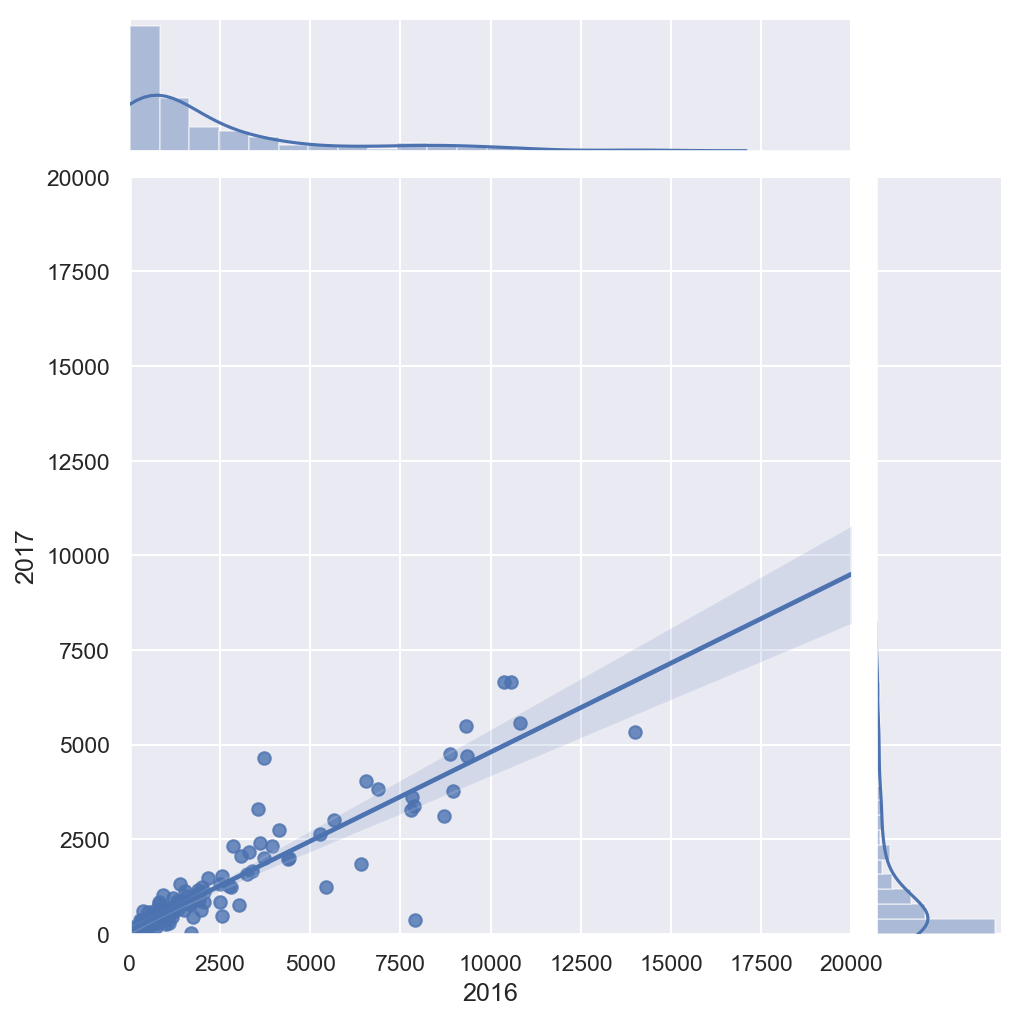}
  \caption{Right-wing}
  \label{fig:10sub2}
\end{subfigure}
\begin{subfigure}{.333\textwidth}
  \centering
  \includegraphics[width=0.9\linewidth]{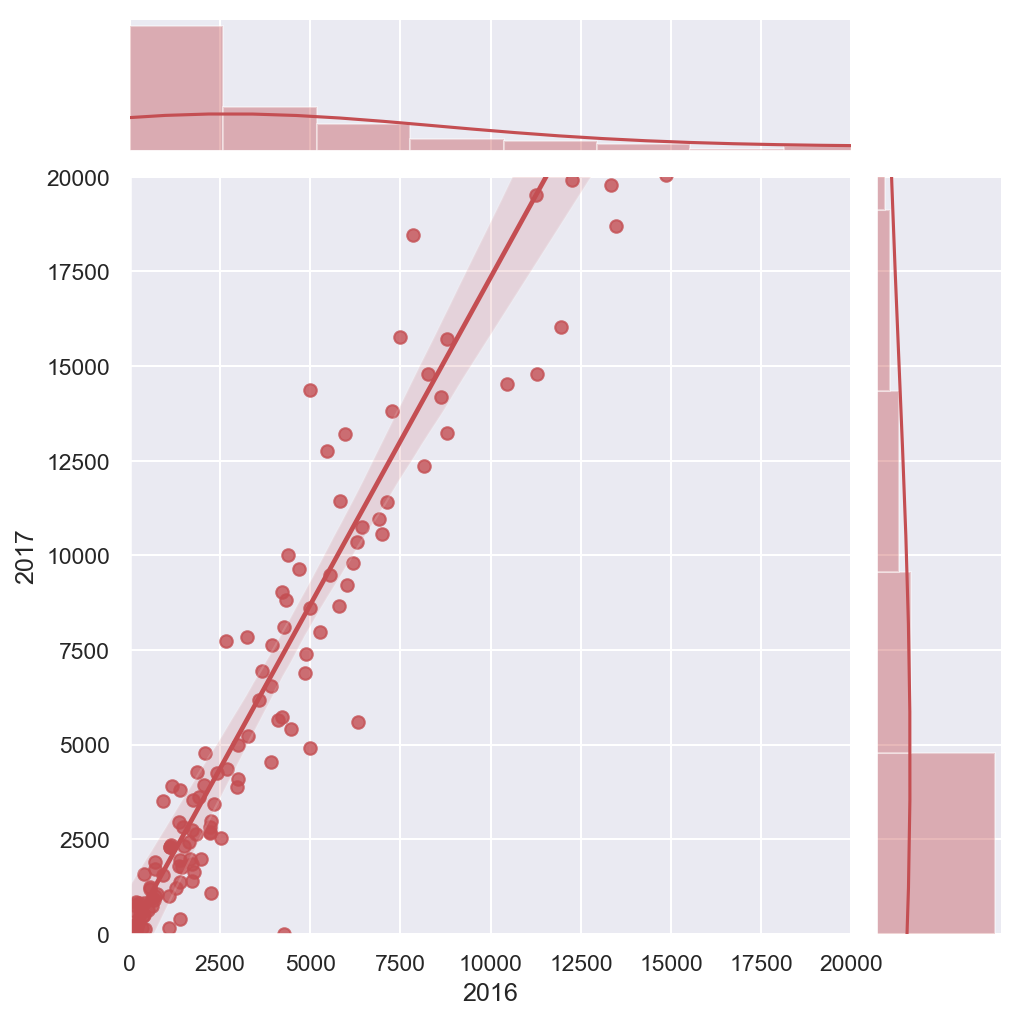}
  \caption{Left-wing}
  \label{fig:10sub3}
\end{subfigure}
\caption{Scatter Plots and Histograms of linear regression with marginal distribution: far-right, right-wing, and left-wing, in 2016 and 2017}
\label{fig:10}
\end{figure}

Figure 10 compares the three political currents, showing different slopes in the diagonal line representing the linear relationship between 2016 and 2017 terms defining narratives. Left-wing saw the frequency of its terms clearly raising during 2017, meaning an increasing of interest with the “Europe” debate. Far-right terms have generally raised; however, this movement is affected only by a few terms whose frequency strongly augmented during 2017. On the contrary, right-wing saw a general decrease of the frequency of its terms, signifying less interest with the “Europe” discussion during 2017. 

\section{Conclusion}

This research allows concluding that narratives in public opinion, understood as social constructs where aggregates of terms are repeated and propagated in society until they are identified as communication patterns, can be characterised through automatic processing and analysis of a corpus. 

An integrative scientific approach allowed processing and analysing the corpus of articles published by the press, including theoretical concepts derived from the social constructivist approach~\cite{8}~\cite{9} and structural linguistics~\cite{10}, and inspired by the mathematical theory of hypernetworks~\cite{11}. A methodological sequence of technologies was applied, including state-of-the-art technologies particularly adapted to the Observatorium corpus, mainly from the domains of Natural Language Processing and Network Theory. New illustrative graphical outputs were proposed, such as the Topic Network Graph, the Narrative Network Graph, and the Narrative Flow Diagram.

The 2016-17 Brexit case study exposes narratives from far-right, right-wing, and left-wing political tendencies, related to Brexit, that were propagated in UK press along this period. The analysis revealed clear distinctions between narratives along the political spectrum. During 2016 Far-right was particularly focused on emigration and refugees. Also, Europe was related to attacks on women and children, sexual offenses, and terrorism. Right-wing was manly focused on internal politics. Left-wing was remarkably mentioning a diversity of non-political topics, such as sports, side by side with economics. During 2017, in general terrorism was less mentioned, and negotiations with EU, namely regarding economics, finance, and Ireland, became central.

\section*{Annexes}

\begin{itemize}
  \item \href{http://theobservatorium.net/brexit/2017-16_Brexit_annexes_71.pdf}{Annex 1 – Topic Keywords}
  \item \href{http://theobservatorium.net/brexit/2017-16_Brexit_annexes_72.pdf}{Annex 2 – Topic Network Graphs }
  \item \href{http://theobservatorium.net/brexit/2017-16_Brexit_annexes_73.zip}{Annex 3 – Summaries }
  \item \href{http://theobservatorium.net/brexit/2017-16_Brexit_annexes_74.pdf}{Annex 4 – Narrative Network Graphs}
  \item \href{http://theobservatorium.net/brexit/2017-16_Brexit_annexes_75.pdf}{Annex 5 – Statistics for Narrative Flow Diagrams}
\end{itemize}

Annexes are deposited at the Observatorium repository:
\href{http://theobservatorium.net/brexit/ }{http://theobservatorium.net/brexit/} and at Dryad: 
\href{https://datadryad.org/review?doi=doi:10.5061/dryad.g3vb582}{https://datadryad.org/review?doi=doi:10.5061/dryad.g3vb582}

\section*{Acknowledgments}

We thank the members of the Complex Systems Digital Campus (\href{http://cs-dc.org}{http://cs-dc.org}), where preliminary versions of this research have been presented and discussed.

\bibliographystyle{unsrt}  


\end{document}